\documentclass[10pt,aps,prl,reprint,superscriptaddress,nofootinbib]{revtex4-1} 
\usepackage{bm,physics}
\usepackage[dvipsnames]{xcolor}
\usepackage{amsfonts,amsmath,amssymb,bm}
\usepackage{graphicx}
\usepackage[utf8]{inputenc}
\usepackage{xspace}
\usepackage{isotope}
\usepackage{xparse}
\usepackage[pdfpagelabels, pdfencoding=auto, psdextra]{hyperref}
\hypersetup{%
  pdfsubject=Paper,
  pdfkeywords={nuclear physics} {Bayesian} {chiral EFT} {reduced basis method} {emulators} {surrogate models} {nucleon deuteron scattering} {Greedy algorithm},
  unicode = true,
  breaklinks = true,
  colorlinks = true,
  linkcolor = blue,
  citecolor = blue,
  menucolor = blue,
  citecolor = blue,
  urlcolor = blue
}

\usepackage{orcidlink}

\graphicspath{{./}} 

\newcommand{\prlsection}[1]{\emph{#1}---}

\renewcommand{\vec}{\mathbf}

\newcommand{\param}{\theta} 
\newcommand{\cE}{c_E}
\newcommand{\cD}{c_D}
\newcommand{\rmat}[2]{\mathcal{R}_{#2 \to #1 } }
\newcommand{\cemul}{c} 
\newcommand{\cemulwa}{c^a} 

\newcommand{\cHH}{b}
\newcommand{\dimL}{N_b}

\newcommand{\dimxi}{N_s}
\newcommand{\dimparam}{N_\theta}
\newcommand{\dimC}{N_c}
\newcommand{\halfplus}{\frac{1}{2}^+}
\newcommand{\halfminus}{\frac{1}{2}^-}
\newcommand{\rfunc}[2]{{\cal F}_{#2 \to #1}}
\newcommand{\schro}{{Schr\"{o}dinger}}

\newcommand{\yvec}{\phi}

\newcommand{\wf}[1]{\Psi^{#1}}
\newcommand{\paramVec}{\vb*{\theta}}  
\newcommand{\thetavec}{\paramVec}

\newcommand{\Amat}{\vb{A}}   
\newcommand{\tAmat}{\widetilde{\Amat}^a}
\newcommand{\svec}{\vb{s}^a}    
\newcommand{\tsvec}{\widetilde{\vb{s}}^a}

\newcommand{\xvec}{\vec{x}^a}

\newcommand{\cvec}{\vb{c} }    
\newcommand{\cvecwa}{\vb{c}^a}

\newcommand{\Xmat}{\vb{X}^a}   
   
\newcommand{\Ymat}{\vb{Y}^a}

\newcommand{\residual}{\vb{r}}
\newcommand{\rvec}{\residual}    
\newcommand{\rvecwa}{\vb{r}^a}

\newcommand{\MeV}{\, \text{MeV}}

\newcommand{\nTwoLO}{N$^2$LO\xspace}
\newcommand{\nThreeLO}{N$^3$LO\xspace}
\newcommand{\chiEFT}{$\chi$EFT\xspace}

\begin{document}

\title{Accurate and Efficient Emulation of Proton-Deuteron Scattering \texorpdfstring{\\}{} via the Reduced Basis Method and Active Learning}

\author{Alex Gnech~\orcidlink{0000-0002-2077-3866}}
\email{agnech@odu.edu}
\affiliation{Department of Physics, \href{https://ror.org/04zjtrb98}{Old Dominion University}, Norfolk, Virginia 23529, USA}
\affiliation{Theory Center, \href{https://ror.org/02vwzrd76}{Jefferson Lab}, Newport News, Virginia 23610, USA}

\author{Xilin Zhang~\orcidlink{0000-0001-9278-5359}} 
 \email{zhangx@frib.msu.edu}
\affiliation{\href{https://ror.org/03r4g9w46}{Facility for Rare Isotope Beams}, \href{https://ror.org/05hs6h993}{Michigan State University}, East Lansing, MI~48824, USA}

\author{Christian Drischler~\orcidlink{0000-0003-1534-6285}}
\email{drischler@ohio.edu}
\affiliation{Department of Physics and Astronomy, \href{https://ror.org/01jr3y717}{Ohio University}, Athens, OH~45701, USA}
\affiliation{\href{https://ror.org/03r4g9w46}{Facility for Rare Isotope Beams}, \href{https://ror.org/05hs6h993}{Michigan State University}, East Lansing, MI~48824, USA}

\author{R.~J. Furnstahl~\orcidlink{0000-0002-3483-333X}}
\email{furnstahl.1@osu.edu}
\affiliation{Department of Physics, \href{https://ror.org/00rs6vg23}{The Ohio State University}, Columbus, OH 43210, USA}

\author{Alessandro Grassi~\orcidlink{0000-0002-5703-3183}}
\email{alessandro.grassi@df.unipi.it}
\affiliation{Department of Physics “E. Fermi”, \href{https://ror.org/03ad39j10}{University of Pisa}, I-56127, Pisa, Italy}
\affiliation{\href{https://ror.org/01vj6ck58}{Istituto Nazionale di Fisica Nucleare, Sezione di Pisa}, I-56127 Pisa, Italy}

\author{Alejandro Kievsky~\orcidlink{0000-0003-4855-6326}}
\email{alejandro.kievsky@pi.infn.it}
\affiliation{\href{https://ror.org/01vj6ck58}{Istituto Nazionale di Fisica Nucleare, Sezione di Pisa}, I-56127 Pisa, Italy}

\author{Laura E. Marcucci~\orcidlink{0000-0003-3387-0590}}
\email{laura.elisa.marcucci@unipi.it}
\affiliation{Department of Physics “E. Fermi”, \href{https://ror.org/03ad39j10}{University of Pisa}, I-56127, Pisa, Italy}
\affiliation{\href{https://ror.org/01vj6ck58}{Istituto Nazionale di Fisica Nucleare, Sezione di Pisa}, I-56127 Pisa, Italy}

\author{Michele Viviani~\orcidlink{0000-0002-4682-4924}}
\email{michele.viviani@pi.infn.it}
\affiliation{\href{https://ror.org/01vj6ck58}{Istituto Nazionale di Fisica Nucleare, Sezione di Pisa}, I-56127 Pisa, Italy}

\date{\today}

\begin{abstract}
We introduce highly accurate and efficient emulators for proton-deuteron scattering below the deuteron breakup threshold. We explore two different reduced-basis method strategies: one based on the Kohn variational principle and another on Galerkin projections of the underlying system of linear equations. We use the adaptive greedy algorithm previously developed for two-body scattering for optimal selection of high-fidelity training points in the input parameter space. We demonstrate that these emulators reproduce \textit{ab~initio} hyperspherical harmonics calculations of $R$-matrix elements with remarkable precision, achieving relative errors as low as $10^{-7}$ with a small number of training points, even in regions of strong nonlinear parameter dependence. They also dramatically accelerate the exploration of the scattering predictions in the parameter space, a capability highly desired for calibrating (chiral) three-nucleon forces against scattering measurements. Our formalism can be further generalized to handle nucleon-deuteron scattering above the breakup threshold. These emulator developments will provide valuable tools to accelerate uncertainty quantification and rigorous parameter inference in the study of nuclear forces. 

\end{abstract}

\maketitle

\prlsection{Introduction}%
Nucleon-deuteron ($Nd$) scattering serves as a fundamental benchmark for understanding nuclear forces, particularly the elusive three-nucleon force (3NF)~\cite{Weinberg:1992yk, Epelbaum:2002vt,Hammer:2012id, Epelbaum:2012zz, Hebeler:2020ocj,Epelbaum:2019kcf,Machleidt:2024bwl}. In chiral effective field theory (\chiEFT) with Weinberg power counting~\cite{Weinberg:1990rz,Weinberg:1991um,Weinberg:1992yk,Epelbaum:2008ga,Machleidt:2011zz}, for example, 3NFs enter at sub-leading order and their strength is characterized by low-energy constants (LECs), including $\cD$ and $\cE$, whose values are not uniquely determined by two-nucleon scattering data \cite{Hebeler:2020ocj}. 
Meanwhile, high-fidelity \textit{ab~initio} scattering calculations  for $Nd$ scattering, performed using methods like the hyperspherical harmonics (HH) approach \cite{Kievsky2008,Marcucci2019} (see also Refs.~\cite{Leidemann:2012hr, Greene:2017cik, Rittenhouse_2011} for the discussions of earlier HH literature), are computationally demanding, due to large bases and complex asymptotics.\footnote{Other types of \textit{ab~initio} $Nd$ scattering calculations have been reviewed,  e.g., in Refs.~\cite{Gloeckle:1995jg,DeltuvaCoulombReview2008,Deltuva:2012kt,Lazauskas:2019rfb}.} This makes systematic exploration of the 3NF parameter space a formidable challenge for applications such as uncertainty quantification and Bayesian inference.

To overcome this bottleneck, 
we develop a computational framework based on the reduced basis method (RBM), a model-driven emulation technique rooted in the field of model order reduction~\cite{hesthaven2015certified,Quarteroni:218966,Duguet:2023wuh,Melendez:2022kid,Drischler:2022ipa}.  Recently, the RBM was rediscovered as eigenvector continuation in quantum physics studies~\cite{Frame:2017fah}. Soon after, such emulators for bound and resonant states~\cite{Frame:2017fah,Sarkar:2020mad,Sarkar:2021fpz,Konig:2019adq,Demol:2019yjt,Ekstrom:2019lss,Demol:2020mzd,Yoshida:2021jbl,Anderson:2022jhq,Giuliani:2022yna,Yapa:2023xyf} and scattering states~\cite{Furnstahl:2020abp,Drischler:2021qoy, Melendez:2021lyq,Zhang:2021jmi,Bai:2021xok, Drischler:2022yfb, Melendez:2022kid, Drischler:2022ipa, Bai:2022hjg, Garcia:2023slj, Odell:2023cun,Liu:2024pqp, Maldonado:2025ftg} were developed. In contrast with data-driven emulations, such as deep learning neural networks and Gaussian processes, our empirical experience is that RBM emulation requires less training data and has higher emulation accuracy, even when extrapolating (see, e.g., Refs.~\cite{Konig:2019adq,Drischler:2022ipa}). Three-body RBM scattering emulators were explored in Ref.~\cite{Zhang:2021jmi} as a precursor of the current study. Here we show that this general approach yields highly accurate and efficient emulators%
\footnote{Also see Ref.~\cite{Witala:2021xqm} for scattering emulators based on certain perturbation expansion in high-fidelity calculations.} for below-threshold proton-deuteron ($pd$) scattering observables. Our framework is directly applicable to neutron-deuteron scattering and can be further developed to handle $Nd$ scattering at higher energies, by properly including the deuteron-breakup channels in both high-fidelity calculations and their emulations.

The core idea of the RBM is to approximate the solution ($\yvec$) of a parameterized problem
by expanding it in an efficient basis, formed by high-fidelity solutions computed at a set of $\dimL$ points in the parameter ($\thetavec$) space:\footnote{Note that $\cemul$ are not the LECs of the underlying interactions.}    
\begin{align}
    \yvec(\thetavec) \approx \sum_{\mu=1}^{\dimL} \cemul_\mu(\thetavec) \yvec(\thetavec_\mu) \label{eq:RBM_ansatz} \ . 
\end{align}
The basis can then be orthornormalized and compressed using the proper orthogonal decomposition (POD)~\cite{Quarteroni:218966}, for better numeric stability~\cite{Giuliani:2022yna, Odell:2023cun, Maldonado:2025ftg}. The expansion leads to a separation of the offline (or training) and online (or evaluation) stages of the emulator: in the offline stage, one obtains $\yvec(\thetavec_\mu)$, called snapshots, and in the online stage, where the emulator is invoked, one solves for $\cemul_\mu(\thetavec)$ at a general emulation point $\thetavec$. The latter is typically achieved through a low-dimensional linear system, resulting from the subspace projection. The large computations are performed at the training stage just once, including the important inputs for constructing the coefficients of the low-dimensional linear algebra equations. 

Specifically, our parameterized problem involves solving a three-body {\schro} equation with a Hamiltonian $H(\thetavec)$ for a scattering state at a given energy $E$. In addition to the application of Eq.~\eqref{eq:RBM_ansatz}, the linear LEC dependence of $H(\thetavec)$ for \chiEFT can be further exploited for computational efficiency in the emulation:%
\footnote{More general affine parameter dependence can be accommodated. Methods exist to approximate non-affine parametric dependence in terms of the affine dependence (see, e.g., Ref.~\cite{Odell:2023cun}).}  
 \begin{equation}
         H(\thetavec) = \sum_{i=0}^{\dimparam}  \theta_i \, H_i \ . \label{eq:Haffinedecomp}
 \end{equation}
Here, $H_0$ is the parameter-independent part of the Hamiltonian, including the kinetic energy, two-body potential, and eventually some fixed three-nucleon potential and the corresponding $\param_{0} = 1$. The other terms $H_i$ are three-body interactions with LECs $\param_{i}$. Note that the subscripts in $\param_{i}$ and  $\thetavec_\mu$ (e.g., in Eq.~\eqref{eq:RBM_ansatz}) are distinct indices. 

Since the solution vector $\yvec(\thetavec)$ has a smooth dependence on $\thetavec$, as demonstrated here and seen in existing studies~\cite{Konig:2019adq,Ekstrom:2019lss,Furnstahl:2020abp,Drischler:2021qoy,Yoshida:2021jbl,Giuliani:2022yna,Garcia:2023slj,Odell:2023cun,Liu:2024pqp}, the scaling of $\dimL$ in terms of $\dimparam$ is expected to be mild~\cite{Sarkar:2020mad,Duguet:2023wuh}, which is one of the key reasons for the high emulation efficiency. The other reason is the affine parameter dependence, such as in the $H(\thetavec)$ matrix in the projected subspace, because at arbitrary $\thetavec$ the matrix can be obtained simply by properly rescaling $H_i$'s matrix elements. 

In this work, we introduce two types of RBM emulators, which differ in how to solve for $\cemul_\mu(\thetavec)$ at the online emulation stage. Both leverage the adaptive power of the so-called greedy algorithm for locally optimal selection of high-fidelity training points $\thetavec_\mu$, which has recently been developed for the quantum scattering problem~\cite{Maldonado:2025ftg}. In addition, this algorithm also mitigates the so-called Kohn anomalies~\cite{PhysRev.124.1468, nesbet1980variational, Drischler:2021qoy,Maldonado:2025ftg}, which can arise in scattering calculations. By leveraging the separation of offline training and online emulation stages, and utilizing the improved selection of training points provided by the greedy algorithm, emulators enable us to capture nontrivial parameter dependencies with minimal computational costs in both training and emulation stages, resulting in rapid predictions with high accuracy. 

The remainder of this paper includes brief discussions of the high-fidelity calculations, emulators, and the numerical demonstrations of their performance. At the end, a summary and outlook are provided. Technical details of this work are provided in the companion paper~\cite{Nd_emulator_2025_long}.
Our developed emulators will be made publicly available on the BUQEYE website, providing users with easy access (e.g., in the form of mini-apps) to these large computations~\cite{Tews:2022yfb}.

\prlsection{HH high-fidelity calculations}%
Our high-fidelity below-breakup $pd$ scattering calculations are performed using the HH method developed by some of us~\cite{Kievsky2008,Marcucci2019}. It is rooted in the Kohn variational principle (KVP)~\cite{Kohn:1948col,JoachainQCT1975,newton2002scattering}, which yields a linear system of equations whose solution provides the scattering wave function. The details can be found, for example, in Refs.~\cite{Kievsky2008,Marcucci2019}. 

Here, the scattering wave function (with a specified total angular momentum $J$ and its projection) is represented as the summation of a spatially localized ``core" and an ``asymptotic''  component, i.e.,
\begin{align}
    |\Psi^{a} \rangle 
    &= \sum_{\xi=1}^{\dimxi} \cHH^{a}_{\xi}|\xi \rangle + \sum_{a'=1}^{\dimC} \left(\delta_{a,a'}  |\Omega_{a'}^R\rangle+\rmat{a'}{a} | \Omega_{a'}^I\rangle\right)  \ , \label{eq:HHwf}
\end{align}
with $a$ denoting the orbital angular momentum $L$ and total spin $S$ of the incoming channel and $a'$ denoting those of all the coupled channels.\footnote{Note that Ref.~\cite{Kohn:1948col} already used the strategy of treating internal and external wave functions separately. Similarly, Eq.~\eqref{eq:HHwf} has been employed in various HH continuum studies~\cite{Kievsky2008,Marcucci2019} (see also early developments in,  e.g., Ref.~\cite{Efros1969HH, Efros:1971vff}).} The core part, expanded in terms of a spatially localized HH basis $|\xi\rangle$, with $\cHH^{a}_\xi$ the expansion parameters, describes the system in the region where the particles are within the range of strong interaction. When $p$ and $d$ are well separated so that only Coulomb interaction remains between them, $|\Psi^{a} \rangle$ approaches its asymptotic limit, as captured by the asymptotic part. 
This part is a linear combination of the regular ($\Omega_{a'}^R$) and irregular ($\Omega_{a'}^I$) solutions of the {\schro} equation---considering the $pd$ relative motion; $\rmat{a'}{a}$, as the relative weight between the $I$ and $R$ components, is the $a \to a '$ $R$-matrix element (or scattering amplitude). Note that all of these components vary with the scattering energy $E$. Moreover, since we work with energies below the deuteron-breakup threshold, the breakup channel is not included in Eq.~\eqref{eq:HHwf}, which, however, becomes relevant for the above-breakup scatterings.  

According to the KVP, the exact $|\wf{a}\rangle$ and $|\wf{a'}\rangle$ wave functions are the stationary solutions of the functional 
\begin{equation}
\rfunc{a'}{a}\big[\wf{a},\wf{a'}\big] \equiv \rmat{a'}{a} -\big\langle\wf{a'}\big| H -E \big| \wf{a}\big\rangle\ . \label{eq:kvp}
\end{equation}
In our high-fidelity calculation of $|\wf{a}\rangle$ with a fixed $a$, the values of $\cHH^a_{\xi}$ and $\rmat{a'}{a}$ ($a'$ indexing all the coupled channels, see Eq.~\eqref{eq:HHwf}) are determined by finding the stationary solution of the diagonal functional $\rfunc{a}{a}$ via varying $\cHH^a_{\xi}$ and $\rmat{a'}{a}$. A system of linear equations for each coupled channel $a$---with a dimension on the   order of $10^3$ to $10^4$ in $pd$ calculations---then follows:
\begin{align}
    \Amat(\paramVec) \xvec(\paramVec) = \svec(\paramVec) \ , \label{eq:general_lineareq}
\end{align}
where $\xvec = \left( \{ \cHH^a_{\xi}(\thetavec)\}_{\xi =1,...,\dimxi},  \{\rmat{a'}{a}(\thetavec) \}_{a' =1,..,\dimC}\right)^\intercal$. The matrix $\Amat$ (which is the same for different $a$ values) and the $\vec{s}^a$ vector are made up of the matrix elements of $H-E$ sandwiched by different components inside the wave functions in Eq.~\eqref{eq:HHwf}; 
more details can be found in the companion paper~\cite{Nd_emulator_2025_long}. We repeat the calculations for all relevant $a$ values and then insert the obtained $|\Psi_a\rangle$ (and $|\Psi_{a'}\rangle$ if $a\neq a'$) into Eq.~(\ref{eq:kvp}) to evaluate  $\rfunc{a'}{a}$ and get  a second-order improved estimate of $\rmat{a'}{a}$.

\prlsection{Emulation Methods}%
Two different emulation strategies are explored and demonstrated in this work.

\textbf{1.\ Variational Emulation:} This approach, following previous scattering emulation studies~\cite{Furnstahl:2020abp, Drischler:2021qoy, Zhang:2021jmi, Garcia:2023slj}, identifies $|\wf{a} \rangle$  (or $|\wf{a} \rangle$ and $|\wf{a'} \rangle$  if $a \neq a'$) as the $ \yvec$ variable in Eq.~\eqref{eq:RBM_ansatz}. It employs the KVP to compute $\cemul_\mu^{a'a}(\thetavec)$ and obtain emulations of the wave functions and $\rmat{a'}{a}$ for a given $(a', a)$ pair  at general $\thetavec$. The degrees of freedom for finding the stationary solution of $\rfunc{a'}{a}$ are $\cemul_\mu^{a'a}$ instead of, e.g., $\cHH^a_\xi$ in Eq.~\eqref{eq:HHwf}.  
The constraint $\sum_{\mu=1}^{\dimL} \cemul_\mu^{a'a}(\thetavec) = 1$, which ensures the correct asymptotic behavior of the ansatz, is implemented by introducing a Lagrangian multiplier $\lambda$~\cite{Furnstahl:2020abp, Drischler:2021qoy, Zhang:2021jmi, Garcia:2023slj}. 
The emulator then operates with a low-dimensional (i.e., $(\dimL+1) \times (\dimL+1)$) linear equation system:
\begin{equation}
    \begin{pmatrix}
       \quad \vb{U}^{a'a}(\thetavec)  \quad &  \vb{1}_{\dimL}\\
       \vb{1}_{\dimL}^T & 0
    \end{pmatrix} 
    \begin{pmatrix}
        \cvec^{a'a}(\thetavec) \\
        \lambda
    \end{pmatrix}=    
    \begin{pmatrix}
        \vb*{\mathcal{R}}_{a\to a'}\\
       1
    \end{pmatrix} \ .   \label{eq:offdiagonal2}
\end{equation}
There are three length-$\dimL$ vectors: $\vb{1}_{\dimL} = \{1\}_{\mu=1\dots \dimL}$, $ \vb*{\mathcal{R}}_{a\to a'} = \{\rmat{a'}{a}{(\thetavec_\mu)}\}_{\mu=1\dots \dimL}$, and $\cvec^{a'a}(\thetavec)=\{\cemul^{a'a}(\thetavec)\}_{\mu=1\dots \dimL}$. The matrix elements of $\vb{U}^{a'a}$, $U^{a'a}_{\mu',\mu} $, is defined as $ \langle \wf{a'}(\thetavec_{\mu'}) | H(\thetavec) - E | \wf{a}(\thetavec_{\mu}) \rangle + (\mu,a \leftrightarrow \mu',a')$ with linear $ \thetavec$ dependence due to Eq.~\eqref{eq:Haffinedecomp}. Following this strategy, the second-order variational improvement to the $R$-matrix elements, which significantly enhances accuracy, can be obtained immediately from Eq.~\eqref{eq:kvp}.

\textbf{2.\ Galerkin-Projection-Based Emulation:}  
Here, we directly emulate the solutions of the linear equation system in Eq.~\eqref{eq:general_lineareq} for a given $a$, by treating $\xvec$ as the $\yvec$ variable in Eq.~\eqref{eq:RBM_ansatz}. That is,  $\xvec(\thetavec)$ is approximated by $\Xmat \cvecwa(\thetavec)$ 
with $\cvecwa$ as the vector notation of $\cemulwa_\mu$ and $\Xmat$ the orthonormalized snapshot matrix with its column vector space spanned by  $\xvec(\paramVec_\mu)$~\cite{Maldonado:2025ftg, Nd_emulator_2025_long}. 

As the next step, our two different emulators use, respectively, Galerkin and Least-Squares Petrov-Galerkin projections to obtain $\cvecwa(\thetavec)$ and construct reduced-order models (G-ROM and LSPG-ROM, respectively)~\cite{Maldonado:2025ftg, Nd_emulator_2025_long}. To introduce these projections, we first define a residual $\rvecwa(\thetavec)$, which has a linear dependence on $\thetavec$:
\begin{eqnarray} \label{eq:residual_affine}
\rvecwa(\paramVec) & \equiv &  \svec - \Amat \Xmat \cvecwa(\paramVec)  \notag \\ 
 & = & \sum_{i=0}^{\dimparam} \left[  \svec_{(i)} - \Amat_{(i)} \Xmat \cvecwa(\paramVec) \right] \param_i \,.
\end{eqnarray} 
Here,  $\Amat_{(i)}$ and $\svec_{(i)}$ are $\thetavec$-independent  (note $\vb*{\theta}_0 = 1$, as in Eq.~\eqref{eq:Haffinedecomp}). The linear (affine) decompositions of $\Amat$ and $\svec$ follows from the same decomposition in Eq.~\eqref{eq:Haffinedecomp}~\cite{Nd_emulator_2025_long}. 

The G-ROM then requires $\rvecwa$ 
to be orthogonal to $\Xmat$'s column vector space 
(i.e., $\xvec(\thetavec_\mu)$), resulting in a low-dimensional linear equation system:
\begin{equation}
    \tAmat(\thetavec)\cvecwa(\thetavec) =  \tsvec(\thetavec) \, ,
\end{equation}
with $\tAmat =  {\Xmat}^\dagger \Amat \Xmat$ 
and $\tsvec = {\Xmat}^\dagger \svec$. The LSPG-ROM  
has the same general form, but its $\tAmat$ and $\tsvec$ are constructed 
as ${\Ymat}^\dagger \Amat \Xmat$ and $ {\Ymat}^\dagger \svec$, with $\Ymat$'s 
column vector space spanned by all the components in 
Eq.~\eqref{eq:residual_affine}---a space larger than the $\Xmat$'s. 
The LSPG-ROM, in principle, offers improved robustness due to its minimization of the norm of $\rvecwa$. 
In both cases,  $\tAmat$ and $\tsvec$ inherit the linear $\thetavec$-dependence from $\Amat$ and $\svec$. A 
detailed description can be found in Refs.~\cite{Maldonado:2025ftg, Nd_emulator_2025_long}.  Afterward, 
the emulated wave functions based on the respective $\cvecwa(\thetavec)$ and $\cvec^{a'}(\thetavec)$ are plugged into the functional of Eq.~\eqref{eq:kvp} to obtain second-order estimates of $\rmat{a'}{a}$.

We use the greedy algorithm from Ref.~\cite{Maldonado:2025ftg} as an adaptive procedure to select training points (i.e., a form of active learning). It starts with a small initial set of high-fidelity training points and the corresponding snapshots. In subsequent iterations, a new training point is strategically chosen where the current emulator estimates the largest error, effectively placing training points where they are most needed. This locally optimal learning procedure in the offline stage aims to minimize the number of expensive, high-fidelity calculations required to achieve a desired accuracy across the parameter space. The norm of the $\rvecwa$ vector ($||\rvecwa(\thetavec)|| $)  serves as a computationally inexpensive proxy for the true error, guiding this iterative selection process, thanks to $\rvecwa$'s linear $\thetavec$-dependence and the separation of offline training and online emulation in its 
evaluations~\cite{Maldonado:2025ftg,Nd_emulator_2025_long}.

\begin{figure*}
    \includegraphics
    {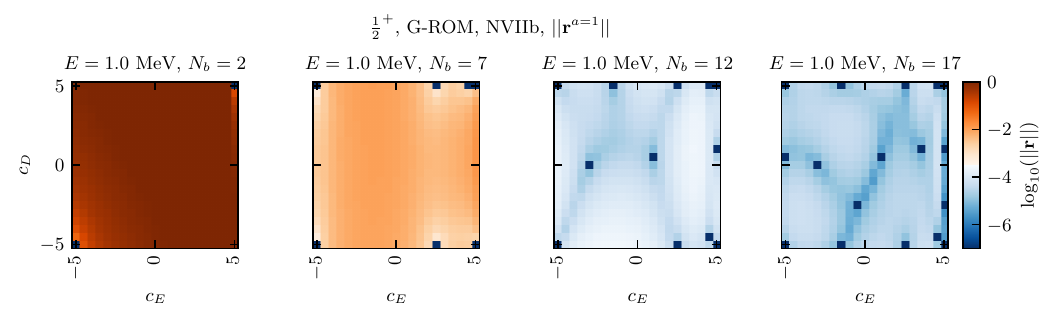}
    \caption{The residuals defined as $||\rvec^{a=1}(\thetavec)||$ for the $\rmat{1}{1}$ estimation for different $(\cE, \cD)$ and scattering energy $E = 1$ MeV, when increasing the number of the training points $\dimL$ sampled using greedy algorithm, for $pd$ scattering ($\halfplus$). The G-ROM emulation and NVIIb force are used. The training points are \emph{effectively} marked by the darkest regions. }
    \label{fig:residuals_GROM_NVIIb_halfplus}
\end{figure*}

\begin{figure*}
    \includegraphics
    {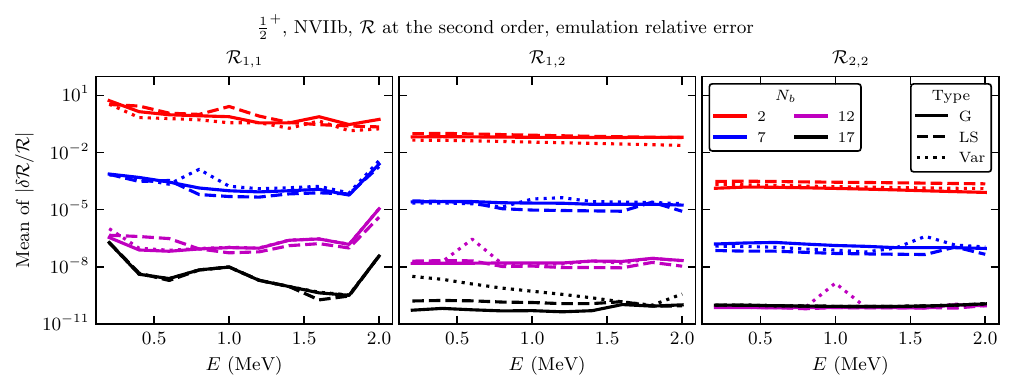}
    \caption{Relative emulation errors for different $R$-matrix elements, averaged over the parameter space as a function of the scattering energy $E$, for the $pd$ scattering ($\halfplus$), using the NVIIb force. ``G'', ``LS'', and ``Var'' in the legend stand for G-ROM, LSPG-ROM, and variational emulation, respectively.}
    \label{fig:emul_rel_error_1dim_comparisions_NVIIb_halfplus}
\end{figure*}

\prlsection{Results}%
We extensively validated our emulators for $pd$ scattering in the $\halfplus$ and $\halfminus$ channels, using the realistic Norfolk two-body chiral \nThreeLO  interactions (NVIIa and NVIIb forces)~\cite{Piarulli2015,Piarulli2016} and \nTwoLO three-body forces as in Ref.~\cite{Baroni2018}. We tested our emulators across a broad range of  $\cE$ and $\cD$ 3NF LECs and scattering energies below deuteron breakup threshold, demonstrating the power of  RBMs coupled with active learning. Here, we focus on the results for $\halfplus$ $pd$ scattering and NVIIb force. A more comprehensive survey of the results of the NVII forces~\cite{Piarulli2015,Piarulli2016} and channels that behave qualitatively similarly can be found in the companion paper~\cite{Nd_emulator_2025_long}. 

Figure~\ref{fig:residuals_GROM_NVIIb_halfplus} exemplifies the greedy algorithm employed in the G-ROM emulations for $pd$ scattering energy $E = 1 \MeV$. Starting from two randomly chosen training points, the greedy algorithm selects the remaining training points successively from a fixed-spaced grid of step size 0.5 in both $c_E$ and $c_D$. When $\dimL\sim 10$, the residual $||\rvec^{a=1}||$ in the worst cases are at a level of $10^{-4}$. Note that we use the $||\rvec^{a=1}||$ from the $\rmat{1}{1}$ calculations as the error proxy, but similar results were obtained using $||\rvec^{a=2}||$. We also checked that $||\rvec^{a=1}||$ as an error proxy and the true error of $\rmat{1}{1}$ follows a similar decreasing trend while the greedy algorithm progresses~\cite{Nd_emulator_2025_long}.  

It is interesting to note that the greedy algorithm adds snapshot calculations at the edges of the parameter space to minimize the errors associated with extrapolating to the boundaries. However, as discussed in Ref.~\cite{Maldonado:2025ftg}, this strategy is not used by the algorithm in higher-dimensional spaces due to the curse of dimensionality. Instead, the algorithm selects a subset of these locations that efficiently reduces the overall emulator error to the requested tolerance.

To assess the performance of our emulators with different $\dimL$ values, we plot in Fig.~\ref{fig:emul_rel_error_1dim_comparisions_NVIIb_halfplus} the averaged relative errors of their emulations on the $c_E$ and $c_D$ grid (as used in Fig.~\ref{fig:residuals_GROM_NVIIb_halfplus}) for various scattering energies.
With only seven training points, the relative errors are already at about $10^{-4}$ or smaller for all the $R$-matrix elements. When $\dimL = 12$, the errors are further down to $10^{-7}$. This shows the power of the proposed emulations. With a few training points, all three emulators can reconstruct the result of the high-fidelity calculation with a high level of precision in a two-dimensional parameter space. 

The G-ROM and LSPG-ROM exhibit similar behavior in this application. The variational emulations, based on the training points identified in the G-ROM greedy algorithm, are close to the other two. This demonstrates a collection of emulator types with similar performance, allowing other developers to choose the one that best suits their individual needs. 

It should be noted that for the chosen energies, $\rmat{1}{1} \approx 1$, while the other matrix elements are $10^2$ to $10^3$ times smaller in magnitude. Therefore, the emulation performance for total cross sections is determined by the $\rmat{1}{1}$'s emulation. However, for observables sensitive to the off-diagonal matrix elements, such as those related to particle polarizations, the off-diagonal emulation errors could be the determining factor. 

The reduction of computational costs offered by these emulators is significant. For each emulation point in the parameter space, the emulator runs by solving a linear system of order $\sim 10$ (taking about 1~$\mu$s), while the high-fidelity calculation deals with a system of order $\sim 10^4$ (taking about 150~s).%
\footnote{The calculations have been performed on a Dell Precision 3660 with  13th Gen Intel(R) Core(TM) i7-13700 CPUs. This speed-up factor is consistent with the expectation based on the computational complexity of solving an $N\times N$ system, which amounts to $\mathcal{O}(N^3)$.} 
The computational speed-up is several orders of magnitude. Of course, generic methods such as splines can be used to interpolate the training points, but the number of interpolating points scales exponentially with the number of parameters $\dimparam$ (e.g., by using a grid of points). In contrast, our emulators require less than 10 training points to achieve sub-percent interpolation accuracy in a two-dimensional space, indicating a much milder scaling of $\dimL$ vs. $\dimparam$, 
which is also seen in bound-state studies working in higher-dimensional parameter space~\cite{Konig:2019adq, Ekstrom:2019lss, Duguet:2023wuh}. This high efficiency is crucial for the practical applications outlined below. Moreover, the memory needed for storing these emulators (essentially small matrices) is a few MBs. Thus, the emulator is highly portable, making it easy for others to access the expensive $pd$ calculations. Note that the emulation performances observed here are similar to those seen in the proof-of-principle study~\cite{Zhang:2021jmi}.

\prlsection{Discussion and Outlook}%
The development of highly accurate and efficient emulators for $pd$ scattering below deuteron breakup threshold represents a significant advancement in nuclear reaction theory and computation. 
It employs scattering emulation technologies~\cite{Furnstahl:2020abp,Drischler:2021qoy, Melendez:2021lyq,Zhang:2021jmi,Bai:2021xok, Drischler:2022yfb, Melendez:2022kid, Drischler:2022ipa, Bai:2022hjg, Garcia:2023slj,Odell:2023cun,Zhang:2021jmi} and greedy algorithms~\cite{Maldonado:2025ftg, Sarkar:2021fpz} to treat the $pd$ system with realistic \chiEFT forces, coupled-channel effects and nontrivial Coulomb interaction. By effectively overcoming the computational bottleneck of high-fidelity calculations, we will be able to apply large-scale Bayesian analyses to constrain 3NF parameters in nuclear forces using experimental scattering data, estimate theoretical uncertainties due to truncating the \chiEFT expansion at finite order, and address recent proposals concerning 3NFs~\cite{Girlanda:2023znc,Cirigliano:2024ocg}.

Both the variational-method-based and Galerkin-projection-based emulators, bolstered by adaptive greedy training, provide powerful tools. Moreover, the second-order variational improvement for the $R$-matrix elements stands out as a key enabler for the unprecedented accuracy achieved.  These emulators can be applied to other computational approaches for solving scattering problems~\cite{Glockle:1983,Gloeckle:1995jg,DeltuvaCoulombReview2008,Deltuva:2012kt,Lazauskas:2019hil,Marcucci:2019hml,Navratil:2016ycn,Navratil:2022lvq,Descouvemont:2010cx,Nielsen:2001hbm} as well as to any kind of Hamiltonian that allows an affine parametrization. Future work will extend these methods to higher scattering energies by incorporating the effects of inelastic channels in emulations, a broader range of nuclear force parameters, and applying them to other complex few-body systems, thereby pushing the boundaries of \textit{ab~initio} nuclear theory. 
It will also be interesting to explore applications of the recently developed Parametric Matrix Models (PMMs)~\cite{Cook:2024toj} to nuclear scattering.\footnote{Other PMM applications can be found, e.g., in Refs.~\cite{Somasundaram:2024zse, Armstrong:2025tza, Yu:2025fyh}.} 
In this respect, our work may provide important guidance for developing active learning approaches for training and identifying optimal outputs of these matrix models (also see Refs.~\cite{Zhang:2024ril,Zhang:2024gac}).
In conclusion, nuclear scattering is a fertile ground for calibrating, validating, and improving nuclear models, facilitated by novel emulator methods such as the one developed in this paper.
Work along these lines is in progress by the DOE STREAMLINE collaboration.

\begin{acknowledgments}
We thank Petar Mlinarić and our STREAMLINE collaborators for fruitful discussions.
We also thank the ECT* for support at the Workshop ``Next generation ab initio nuclear theory'' during which this work was presented and further developed.
This material is based upon work supported by the U.S. Department of Energy, Office of Science, Office of Nuclear Physics, under the FRIB Theory Alliance award DE-SC0013617, under the STREAMLINE collaboration awards DE-SC0024586 (Michigan State University),  DE-SC0024233 (Ohio University) and DE-SC0024509 (Ohio State University), the STREAMLINE 2 collaboration award DE-SC0026198, and by the National Science Foundation under awards PHY-2209442/PHY-2514765.
R.J.F. also acknowledges support from the ExtreMe Matter Institute EMMI at the GSI Helmholtzzentrum für Schwerionenforschung GmbH, Darmstadt, Germany.
The work of A.G. (A.~Gnech) is supported by the U.S. Department of Energy through the Nuclear Theory for New Physics Topical Collaboration,  under contract DE-SC0023663. A.G. (A.~Gnech) acknowledges also the support of Jefferson Lab supported by the U.S. Department of Energy under contract DE-AC05-06OR23177.
A.G. (A.~Grassi), A.K., L.E.M. and M.V. acknowledge the financial support of the
European Union - Next Generation EU, Mission 4 Component 1, CUP
I53D23001060006.
The following open-source Python libraries were used to generate the results in this work:
\texttt{matplotlib}~\cite{Hunter:2007},
\texttt{numpy}~\cite{harris2020array}, and
\texttt{scipy}~\cite{2020SciPy-NMeth}.

\end{acknowledgments}

%

\end{document}